\documentclass[useAMS,usenatbib,referee]{mn2e}
\usepackage{graphicx}

\title[The puzzling abundance pattern of HD134439 and HD134440]
{The puzzling abundance pattern of HD134439 and
HD134440\thanks{Based on observations made with ESO Telescopes at
the La Silla or Paranal Observatories from ESO archive 
and data collected at Subaru Telescope
and obtained from the SMOKA science archive at Astronomical Data
Analysis Center, which are operated by the National Astronomical
Observatory of Japan.}}

\author[Y.Q. Chen and G. Zhao]
{Y.Q. Chen and G. Zhao\thanks{E-mail:gzhao@bao.ac.cn}\\
National Astronomical Observatories, Chinese Academy of Sciences, A20, Datun Road, Chaoyang District, 100012 Beijing, P.R.China}

\begin{document}
\newcommand{\afe}{\mbox{\rm [$\alpha$/Fe]}}
\newcommand{\feh}{\mbox{\rm [Fe/H]}}
\newcommand{\ffe}  {\mbox{${\rm [\frac{Fe}{H}]}$}}
\newcommand{\teff}{$T_{\rm eff}$}
\newcommand{\logg}{$\log g$}
\newcommand{\kmprs}  {\mbox{\rm\,km\,s$^{-1}$}}
\newcommand{\NaFe} {\mbox{\rm [Na/Fe]}}
\newcommand{\MgFe} {\mbox{\rm [Mg/Fe]}}
\newcommand{\AlFe} {\mbox{\rm [Al/Fe]}}
\newcommand{\SiFe} {\mbox{\rm [Si/Fe]}}
\newcommand{\ZnFe} {\mbox{\rm [Zn/Fe]}}
\newcommand{\KFe} {\mbox{\rm [K/Fe]}}
\newcommand{\CaFe} {\mbox{\rm [Ca/Fe]}}
\newcommand{\ScFe} {\mbox{\rm [Sc/Fe]}}
\newcommand{\TiFe} {\mbox{\rm [Ti/Fe]}}
\newcommand{\VFe} {\mbox{\rm [V/Fe]}}
\newcommand{\CrFe} {\mbox{\rm [Cr/Fe]}}
\newcommand{\MnFe} {\mbox{\rm [Mn/Fe]}}
\newcommand{\CoFe} {\mbox{\rm [Co/Fe]}}
\newcommand{\CuFe} {\mbox{\rm [Cu/Fe]}}
\newcommand{\NiFe} {\mbox{\rm [Ni/Fe]}}
\newcommand{\BaFe} {\mbox{\rm [Ba/Fe]}}
\newcommand{\fehta} {\mbox{\rm [FeI/H]}}
\newcommand{\feht} {\mbox{[\rm FeII/H]}}
\newcommand{\YFet} {\mbox{[\rm YII/FeII]}}
\newcommand{\SiFet} {\mbox{[\rm SiI/FeII]}}
\newcommand{\ZnFet} {\mbox{[\rm ZnI/FeII]}}
\newcommand{\ScFet}{\mbox{[\rm ScII/FeII]}}
\newcommand{\BaFet}{\mbox{[\rm BaII/FeII]}}

\date{Accepted ....; Received ...}

\pagerange{\pageref{firstpage}--\pageref{lastpage}} \pubyear{2005}

\maketitle

\label{firstpage}

\begin{abstract}
Abundances of 18 elements are determined for the common
proper-motion pair, HD\,134439 and HD\,134440, which shows high
[Mn/Fe] and low [$\alpha$/Fe] ratios as compared to normal halo
stars. Moreover, puzzling abundances are indicated from elements
whose origins are normally considered to be from the same
nucleosynthesis history. Particularly, we have found that [Mg/Fe]
and [Si/Fe] are lower than [Ca/Fe] and [Ti/Fe] by 0.1-0.3 dex.
When elemental abundances are interpreted in term of their
condensation temperatures ($T_C$), obvious trends of [X/Fe] vs.
$T_C$ for $\alpha$ elements and probably iron-peak elements as
well are shown. The hypothesis that these stars have formed from a
dusty environment in dSph galaxy provides a solution to the
puzzling abundance pattern.

\end{abstract}

\section{Introduction}
It is well accepted that metal-poor halo stars have the enhanced [$\alpha$/Fe]
\footnote{Here $\alpha$ usually denotes
Mg, Si, Ca and Ti elements.}
ratio of 0.4 dex
while this ratio
decreases with increasing metallicity for stars with $\feh>-1$.
In contrary to this traditional view, a few metal-poor stars showing
relatively low [$\alpha$/Fe] in the metallicity range of $-2.0 <\feh<-0.7$
have been reported \citep{Nissen97,Ivans03}, and abundances and kinematics accumulated in spectroscopic observations
indicate that
some of these low-$\alpha$ stars could be accreted from nearby dwarf
galaxies  which have a lower star formation rate than the Galactic
halo \citep{Gilmore98}.

Recently, \citet{Shigeyama03}
have proposed a new possible origin for low-$\alpha$ stars in the
metallicity of $-2.0 <\feh<-1.0$. They suggested that some metal-poor
stars harbor planetary systems and show low-$\alpha$ ratios
due to engulfing planetesimals. Inspired by this conjecture, we attempt to
investigate the abundance pattern of HD\,134439
and HD\,134440, which
are the best test samples with the reported low [$\alpha$/Fe]
\citep{King97}
and the metallicity in the middle of $-2.0 <\feh<-1.0$.
\citet{Ivans03}
's sample of low-$\alpha$ stars is close to $\feh =-2.0$ while
most low-$\alpha$ stars
in \citet{Nissen97}
have $\feh > -1.0$.
Moreover, these two stars are a common proper-motion pair
and abundance difference between this pair provides a way
to inspect the possibility of harboring planetary systems.

Two previous works on abundance determination of this pair
\citep{King97,Fulbright00} gave inconsistent results for some
important elements. The abundance ratios of [Na/Fe], [Mg/Fe] and
[Si/Fe] show deviations of 0.15-0.25 dex. Since these ratios
provide crucial information on the formation environment of this
pair, it is desirable to make a new analysis with high quality
spectra from UVES/VLT and HDS/SUBARU by an updated analysis method
in the present work. {\bf Moreover, our work will derive
abundances for more elements based on more lines. In particular},
abundances of a few interesting elements, such as K, Sc, Mn, Cu,
Co and Zn, are not available in the literature. The abundance
ratios of these elements provide key clues to their origin and
give new information on their nucleosynthesis history.

\section{Observations and Data Reductions}
In order to reduce analysis error and to facilitate the
interpretation of resulting abundance, we select HD\,211998 as a
standard star for comparison. HD\,211998 is a subgiant, which is expected to be
immune to the planet (if existed) formation process  according to
\citet{Shigeyama03}, and its metallicity is the same as our target stars.

High resolution UVES spectra were retrieved from the ESO archive
\footnote{http://www.eso.org} with the resolving power of 40,000
and the signal-to-noise ratio of 200 for HD\,134439/40 and
R$\sim$60\,000 and S/N$\sim$400 for HD\,211998. Subaru archive
data \footnote{http://smoka.nao.ac.jp} \citep{Baba02} supplied
additional HDS \citep{Noguchi02} spectra for HD\,134439 with
R$\sim$90\,000 and S/N$\sim$300. {\bf Note that the resolution and
signal-to-noise of our HDS spectra for HD\,134439 are higher than
those of \citet{King97} and \citet{Fulbright00}, while UVES
spectra have similar resolution and signal-to-noise ratio as those
of \citet{King97}. In comparison with \citet{Fulbright00}, the
signal-to-noise of our UVES spectra is significantly higher.}

The spectra were reduced by using standard routines in MIDAS
software for order identification, background subtraction,
flat-field correction, order extraction and wavelength
calibration. Bias, dark current and scattered light corrections
are included in the background subtraction. The spectra were then
normalized by a continuum function determined by fitting a spline
curve to a set of pre-selected continuum windows estimated from
the solar atlas. Finally, correction of radial velocity shift,
measured from 20 lines, was applied and the equivalent widths were
measured by Gaussian fitting.

Fig.~1 shows the comparison of equivalent widths between this work
and \citet{King97}. Note that two points showing significant
deviations as large as 15 m\AA\ are Mg lines at $\lambda$5711\AA\
for HD\,134439 and HD\,13440 respectively.  We have checked that
our spectra of HD\,134439 from UVES and HDS give the same
equivalent widths for this line. Since our HDS spectra have very
high quality, and the agreement indicates that our values for this
line are reliable. The comparison of equivalent widths of
HD\,134439 from our UVES and HDS spectra for 189 lines in common
in Fig.~2 also gives consistent values with a deviation of
$\Delta{\rm EW(HDS-UVES)}=-2.01 \pm 3.18$ m\AA.

\begin{figure}
\includegraphics[width=84mm]{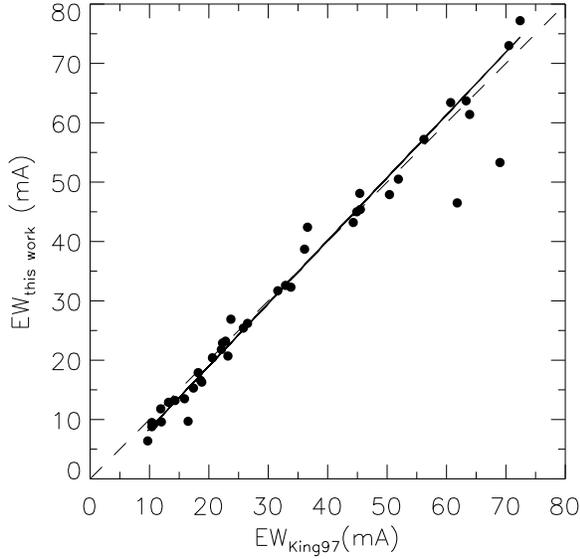}
\caption{The comparison of equivalent widths between this work
and \citet{King97}. The thick line is the linear fit to the points,
{whereas the dashed line is the one-to-one relation.}}
\label{fig:ewcom}
\end{figure}

\begin{figure}
\includegraphics[width=84mm]{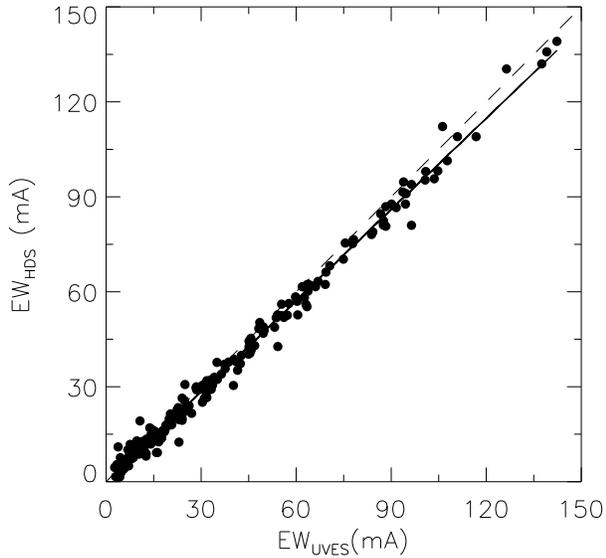}
\caption{The comparison of equivalent widths of HD\,134439 from our
UVES and HDS spectra.
The thick line is the linear fit to the points, {whereas the dashed line is the one-to-one relation.}}
\label{fig:ewcom2}
\end{figure}

\section{Abundance Analysis}

\begin{table}
\caption{Atmosphere parameters and [X/Fe] ratios for HD\,134439(1), HD\,134440(2) and HD\,2119988(3). X indicates different elements.}
\label{tb:abu}
\setlength{\tabcolsep}{0.03cm}
\begin{tabular}{lrrrcrrr}
\noalign{\smallskip}
\hline
 $[{\rm X/Fe]}$& (1) & (2) & (3) \vline & $[{\rm X/Fe}]$& (1) & (2) & (3)\\
\hline
\teff (K)  & 5021& 4852   & 5241 \vline&Sc & $-$0.03 & $-$0.04 & 0.01\\
\logg  & 4.66& 4.64& 3.34 \vline&Ti &  0.18 & 0.20  & 0.34 \\
$\xi_t$     & 1.10 & 1.10 & 1.10 \vline&V  &  0.01 &0.05 &  $-$0.07 \\
FeI &  $-$1.43& $-$1.45 &$-$1.45 \vline& Cr & $-$0.01 &0.06 &$-$0.07\\
FeII &$-$1.43   & $-$1.45  & $-$1.48 \vline&Mn&  $-$0.41&$-$0.37 &$-$0.59\\
Na  & $-$0.48 &$-$0.49 & $-$0.13 \vline& Co & $-$0.03 &  0.05 & $-$0.03 \\
Mg    & $-$0.09 &$-$0.10 & 0.22 \vline& Ni &$-$0.12&$-$0.13& $-$0.03\\
Al   &  $-$0.22  &$-$0.19 &$-$0.12 \vline&Cu & $-$0.69 & --  &  $-$0.63 \\
Si   & 0.04 & 0.04 &0.33 \vline&Zn & $-$0.04&$-$0.06& 0.02\\
K    &  0.10 & 0.08 & 0.77 \vline& Y & $-$0.22 & $-$0.28 & $-$0.03\\
Ca   &  0.09 & 0.08 & 0.36 \vline& Ba & $-$0.35  & $-$0.36 & $-$0.04 \\
\hline
\end{tabular}
\end{table}

\subsection{Stellar Parameters}
The effective temperatures are based on \citet{Alonso96} 's IRFM
\teff\ scale and (b-y) color indices. Reddening is assumed to be
negligible since all stars are within 30\, pc from the Sun. We
have checked that the slopes of iron abundances versus excitation
potential are small. Gravities are derived from Hipparcos
parallaxes \citep{Chen00} and metallicities are iterated by
abundance analysis. Microturbulences are determined by forcing FeI
lines with different strengths to give consistent abundances.
Stellar parameters and resulting abundances are presented in Table
1.

Our temperatures are generally consistent with those in
\citet{King97} with deviations of 21\,K and 67\,K for HD\,134439
and HD\,134440, respectively, but they are significantly higher
than those in \cite{Fulbright00}. As noticed in \citet{Chen06},
spectroscopically-derived temperatures for solar type stars show
systematically deviations in opposite directions in different
works, and they strongly depend on stellar model atmosphere, the
selection of lines, the adopted atomic data, and so on. We thus
prefer the IRFM-based photometric temperatures. Furthermore,
surface gravities in the present work are derived from parallaxes
which are more reliable than those in \citet{King97} and
\citet{Fulbright00}. Finally, we note that the microturbulences of this pair
adopted by \citet{Fulbright00} are 0.7 $\kmprs$ which are quite low
as compared with the usually adopted value of $1.5 \kmprs$ for
metal-poor stars in the literature.

The internal error in \teff\ derived in the present work is
around 70\,K as shown in \cite{Nissen04}. Considering the
significant discrepancy with spectroscopically-derived value in
\cite{Fulbright00}, we adopted the error of 150 K in \teff\ as
the maximum value of temperature error.
Errors of other atmospheric parameters are estimated
to be 0.15 dex in logg, 0.1 dex in $\feh$
and 0.3 $\kmprs$ in $\xi_t$.

\subsection{Abundances and Error Analysis}
The unblended lines with strength strong enough for measurement
were carefully selected by using the solar atlas of
\citet{Moore66}. The oscillator strengths are mainly the same or
taken from the same sources as those in \citet{Chen00}, which are
checked for consistency with the Sun being one of "standard"
stars. The atomic line data and equivalent widths for HD\,134439,
HD\,134440 and HD\,211998 are presented in Table A1
\footnote{Table A1 will be available in electronic form at the CDS
via anonymous ftp to cdsarc.u-strasbg.fr (130.79.128.5)}, which is
published electronically only.

The model atmospheres were interpolated from a grid of
plane-parallel, LTE models provided by \citet{Kurucz95} in which
convective overshoot is switched off. The ABONTEST8 program,
developed by P. Magain in the Li\'ege group, was used to carry out
the calculations of theoretical equivalent widths of lines and
abundance was derived by matching the theoretical equivalent
widths to the observed values. Hyperfine structures for Sc, Mn, Cu
and Ba are checked to have small effects since the adopted lines
are weak in metal-poor stars. Solar abundances from
\citet{Grevesse98} are adopted to derive the relative abundances.
{\bf Since Si and Zn lines corresponds to high excitation levels
and are weak in our stars, they are formed from deep in the
stellar atmospheres which is similar as that of FeII lines. In
view of this, iron abundances from FeII lines were adopted to derive [Si/Fe]
and [Zn/Fe] ratios. On the contrary, iron abundances
from FeI lines were used to obtain [Ba/Fe] ratio because BaII
lines behave more like FeI lines rather than FeII lines.} In this
way, we can expect that errors of [X/Fe] ratios can be greatly
reduced. For other elements, we follow the general rule that
elemental abundances derived from neutron lines are relative to
FeI lines and abundances from ionized lines are relative to FeII
lines.

\begin{figure}
\includegraphics[width=84mm]{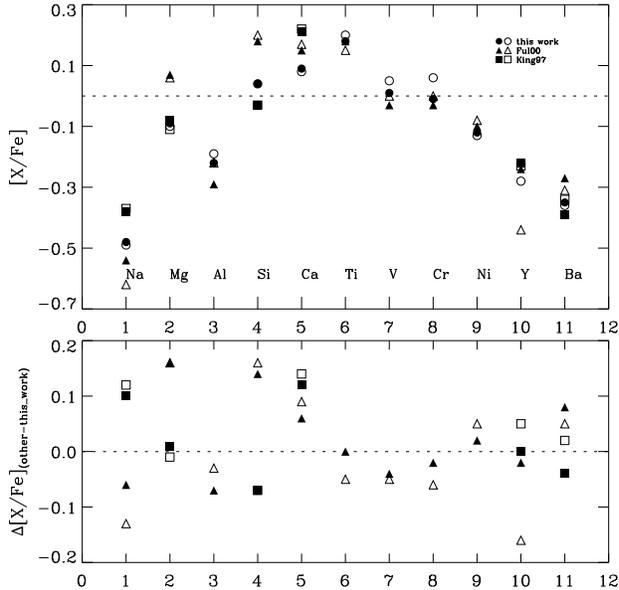}
\caption{The comparison of abundances between this work (circle)
and \citet{King97} (triangle), and \citet{Fulbright00} (square) for
HD134439 (filled symbols) and HD\,134440 (open symbols).
The low panel shows the deviation of abundances between
this work and \citet{King97} and \citet{Fulbright00} for the two stars.}
\label{fig:ewabu}
\end{figure}

The uncertainties estimated by a change of 150\,K in \teff, 0.2
dex in \logg, 0.1 dex in \feh\ and 0.3 \kmprs in $\xi_t$ are shown
in Table 2, where numbers with brackets indicate the internal
errors by using 70\,K as an error in temperature. It shows that
abundance errors are 0.12 dex for iron abundances and less than
0.1 dex for [X/Fe] ratios even the external error of 150\,K is
used. Clearly, main errors in abundances of most elements come
from the uncertainty of temperature. The abundance errors will be
greatly reduced (about 0.07 dex) when the internal error in
temperature of 70\,K is adopted.

The comparisons of uur abundances with those in \citet{King97} and
\citet{Fulbright00} are shown in Fig.~3. Generally, our values are
quite similar with \citet{King97}. The [Na/Fe] and [Ca/Fe] show
the largest deviations around 0.1 dex. Note that more lines are
used in the present work to derive abundances while only one line
for Na and three lines for Ca are adopted in \citet{King97}. In
comparison with Fulbright (2000), there is a large disagreement by
0.2-0.3 dex in [Na/Fe] and [Si/Fe]. Since our abundances are
derived with a more reasonable atmospheric parameters based on
spectra with higher signal-to-noise, they may be more reliable.

\begin{table*}
\setlength{\tabcolsep}{0.05cm}
\caption[]{Abundance errors for HD\,34439 and HD\,134440.}
\label{tb:abuerr}
\begin{tabular}{lrrrrrrrrrrrrrrrr}
\noalign{\smallskip}
\hline
\noalign{\smallskip}
 & $\frac{\sigma_{\rm EW}}{\sqrt{\rm N}}$ & $\Delta$ \teff & ($\Delta$ \teff) & $\Delta$ \logg & $\Delta$ \ffe   & $\Delta \xi_t$ &  $\sigma_{\rm tot}$ &  $\sigma_{\rm tot}$   & $\frac{\sigma_{\rm EW}}{\sqrt{\rm N}}$&$\Delta$ \teff &($\Delta$ \teff) &$\Delta$ \logg & $\Delta$ \ffe  & $\Delta \xi_t$ & $\sigma_{\rm tot}$ & $\sigma_{\rm tot}$\\
       &     & +150K &(+70K)  & +0.2  & +0.1  & +0.3  & 150K & (70 K) & &   +150K & (+70K)  & +0.2  & +0.1  & +0.3  &150K & (70 K) \\[-1.8mm]
 & \multicolumn{8}{c}{-------------------------------------------------------------------------} & \multicolumn{8}{c}{-------------------------------------------------------------------------} \\[-2.0mm]
& \multicolumn{8}{c}{HD134439:5021~K/4.7~/-1.4/1.1}  & \multicolumn{8}{c}{HD134440:4852~K/4.6~/-1.4/1.1} \\
\noalign{\smallskip}
\hline
\noalign{\smallskip}
$\Delta \fehta$& 0.009& 0.122&(0.055)& $-.$007& 0.010& $-.$016& 0.129 &(0.059)& 0.009& 0.111&(0.056)& 0.005& $-.$003& $-.$016& 0.118&(0.059)\\
$\Delta \NaFe$& 0.056& 0.038&(0.016)& $-.$049& $-.$006& 0.012& 0.084&(0.077)& 0.038& 0.040&(0.028)& $-.$057& $-.$011& 0.013& 0.081&(0.075)\\
$\Delta \MgFe$& 0.044& $-.$046&($-.$032)& $-.$001& $-.$018& 0.013& 0.067&(0.058)& 0.046& $-.$035&($-.$022)& 0.000& 0.002& 0.014& 0.059&(0.052)\\
$\Delta \AlFe$& 0.014& 0.066&($-.$025)& 0.012& $-.$016& 0.016& 0.072&(0.037)& 0.031& $-.$033&($-.$014)& 0.005& $-.$005& 0.015& 0.047&(0.037)\\
$\Delta \SiFe$& 0.043& $-.$115&($-.$065)& 0.050& $-.$005& 0.015& 0.134&(0.094)& 0.033& $-.$120&($-.$073)& 0.059& 0.029& 0.015& 0.141&(0.105)\\
$\Delta \KFe$& 0.026& 0.071&(0.055)& $-.$023& 0.051& $-.$011& 0.094&(0.083)& 0.019& 0.070&(0.050)& $-.$102& $-.$043& $-.$009& 0.132&(0.123)\\
$\Delta \CaFe$& 0.021& 0.014&(0.010)& $-.$031& $-.$011& 0.004& 0.041&(0.040)& 0.024& 0.038&(0.028)& $-.$048& $-.$023& 0.004& 0.069&(0.064)\\
$\Delta \TiFe$& 0.034& 0.097&(0.047)& $-.$016& $-.$012& $-.$005& 0.105&(0.061)& 0.040& 0.125&(0.066)& $-.$030& $-.$040& $-.$009& 0.141&(0.092)\\
$\Delta \VFe$& 0.048& 0.080&(0.041)& 0.005& $-.$018& 0.010& 0.095&(0.066)& 0.039& 0.116&(0.058)& 0.002& $-.$018& 0.006& 0.123&(0.072)\\
$\Delta \CrFe$& 0.036& 0.057&(0.030)& $-.$030& $-.$009& $-.$007& 0.074&(0.056)& 0.041& 0.080&(0.045)& $-.$043& $-.$033& $-.$006& 0.105&(0.081)\\
$\Delta \MnFe$& 0.030& 0.038&(0.020)& 0.010& $-.$014& 0.010& 0.051&(0.040)& 0.045& 0.069&(0.033)& 0.011& $-.$006& 0.005& 0.083&(0.057)\\
$\Delta \CoFe$& 0.030& $-.$018&($-.$008)& 0.037& $-.$005& 0.012& 0.052&(0.049)& 0.045& $-.$011&($-.$010)& 0.052& 0.022& 0.011& 0.073&(0.074)\\
$\Delta \NiFe$& 0.018& $-.$040&($-.$014)& 0.039& 0.002& 0.011& 0.059&(0.046)& 0.018& $-.$031&($-.$024)& 0.054& 0.028& 0.012& 0.071&(0.068)\\
$\Delta \CuFe$& 0.009& $-.$034&($-.$016)& 0.043& $-.$003& 0.015& 0.057&(0.048)& --   &  -- & -- &  --  & --   & --   & -- & -- \\
$\Delta \ZnFe$& 0.015& $-.$134&($-.$062)& 0.062& 0.005& 0.008& 0.149&(0.090)& 0.015& $-.$121&($-.$070)& 0.070& 0.028& 0.009& 0.144&(0.104)\\
$\Delta \BaFe$& 0.025& $-.$066&($-.$029)& 0.062& 0.024& $-.$004& 0.097&(0.077)& 0.058& $-.$034&($-.$025)& 0.071& 0.032& $-.$003& 0.103&(0.100)\\
\noalign{\smallskip}
\hline
\hline
\noalign{\smallskip}
$\Delta \feht$& 0.051& -.056&(-.029)& 0.081& 0.026& -.011& 0.105 &(0.104)& 0.060& -.047&(-.038)& 0.113& 0.046& -.009& 0.133&(0.130)\\
$\Delta \SiFet$& 0.043& 0.063&(0.021)& -.038& -.021& 0.010& 0.102&(0.082)& 0.033& 0.038&(0.021)& -.049& -.020& 0.008& 0.095&(0.089)\\
$\Delta \ScFet$& 0.031& 0.070&(0.044)& -.003& 0.000& 0.007& 0.092&(0.070)& 0.038& 0.082&(0.044)& -.009& 0.001& 0.005& 0.109&(0.084)\\
$\Delta \ZnFet$& 0.015& 0.044&(0.024)& -.026& -.011& 0.003& 0.074&(0.064)& 0.015& 0.037&(0.024)& -.038& -.021& 0.002& 0.084&(0.079)\\
$\Delta \YFet$& 0.015& 0.088&(0.053)& -.003& -.001& 0.005& 0.103&(0.069)& 0.015& 0.098&(0.053)& -.011& -.007& 0.000& 0.117&(0.082)\\
$\Delta \BaFet$& 0.025& 0.112&(0.069)& -.026& 0.008& -.009& 0.129&(0.084)& 0.058& 0.124&(0.069)& -.037& -.017& -.010& 0.155&(0.116)\\
\noalign{\smallskip}
\hline
\end{tabular}
\end{table*}

\section{Abundances and Discussions}

The common proper-motion pair is presumed to be formed out from
the same cloud, and thus we can expect that HD\,134439 and
HD\,134440 have the same age and chemical history. Unfortunately,
it is difficult to determine their ages since they are on the main
sequence. For chemical composition, identical abundances between
these two stars have been found by \citet{King97} for six elements
and are confirmed in the present work for more elements as shown
in Fig.~4. One exception is the lithium abundance. We obtained the
abundance of 0.48 dex for HD\,134439 which is close to the value
(0.53 dex) in \citet{King97}. The LiI line at $\lambda$6708\AA\ is
undetected for HD\,134440 and an upper limit of Li abundance is
0.06 dex based on the assumed equivalent width of 3.0 m\AA.

Some common proper-motion pairs, e.g. 16\,Cyg
and HD\,219542,
are reported to harbor planets and abundance differences
in planet host stars have been suggested \citep{Laws01,Gonzalez01,Sadakane03}.
However, \citet{Desidera04} recently studied 23 common proper
motion pairs with solar metallicity, some of which harbor planets,
and suggested that differences in iron abundances between the pairs
are less than 0.07 dex. Presently, it is premature to exclude the possibility
of the presence of planets on the HD\,134439/40
pair based on their identical abundances.

\begin{figure}
\includegraphics[width=84mm]{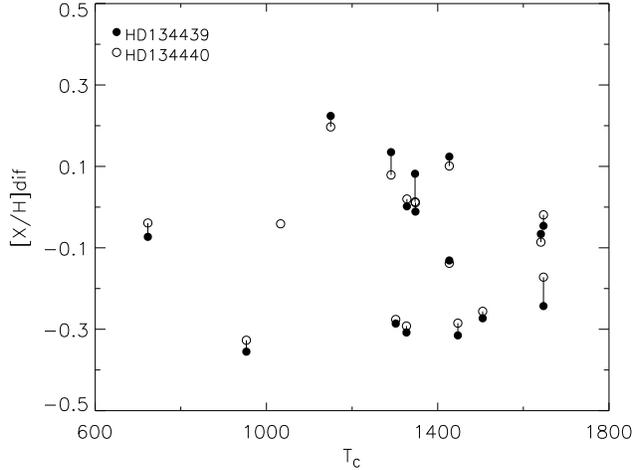}
\caption{The [X/H] (differential to  HD\,211998)  vs. $T_{C}$ for
HD\,134439 (filled circles) and HD\,134440 (open circles)
where X indicates different elements and thick lines
connect abundances of the two stars for the same element.}
\label{fig:AbuTc}
\end{figure}

As reported by \citet{King97} and confirmed in the present work,
this pair has lower $\afe$ ratio than normal stars at $\feh \sim
-1.5$. According to \citet{Shigeyama03}, low-$\alpha$ metal-poor
stars might harbor planets and they accreted planetesimals which
enhanced the surface content of Fe and thus reduced the $\afe$
ratios. In this regard, we would expect an increasing enhancement
of elements with higher condensation temperatures (hereafter,
$T_C$). Simply, we adopted the condensation temperature of the
solar system from \citet{Lodders03} since there is no better
source in the literature. However, there is no any clear trend in
Fig.~4 where abundances, relative to the comparison star
HD\,211998, versus  $T_C$ are shown for both HD\,134439 and
HD\,134440. Especially, Zn has low $T_C$ but does not show
underabundant [X/Fe] ratio as much as Mg and Si elements which
have higher $T_C$. Moreover, [Mn/Fe] is higher than HD\,211998
which is contrary to the prediction by \citet{Shigeyama03} that
low $\afe$ stars with low [Mn/Fe] harbor planetary systems. {\bf
In addition, our abundances of Mn, Co, Cr and Ni elements, 
as presented in Table 1, do not support
the suggestion by \citet{Tsujimoto06} that HD\,134439/40 are
relics of a population III SNIIIa in which they predicted the
deficiency of odd-number elements, Mn and Co, relative to
even-number elements, Cr and Ni.}

HD\,134439/40 have the same and special kinematic parameters
($R_{apo}$=45\,kpc, $V_{LSR}$=-315\,\kmprs, $Z_{max}$=1.7\,kpc)
which are thought to be a evidence of accreted substructure in the
halo field according to \citet{Carney96}. Similarly, other eight
low $\afe$ stars found by \citet{Nissen97} also show large
$R_{apo}$ and $Z_{max}$ and they were related to the accretion of
our Galaxy from nearby dwarf spheroidal (dSph) galaxies. The
inspection of our resulting abundances of HD\,134439/40 with stars
in dSph galaxies by \citet{Shetrone01,Shetrone03} shows generally
similar abundance pattern. For example, the [Na/Fe] ratio in our
stars is lower than [Al/Fe], which is consistent with the Fig. 2
of Shetrone et al. (2003) for stars in dSph galaxies. Moreover,
the [Ni/Fe] of this pair is lower than that of the reference star
HD\,211998 by 0.1 dex, indicating a Na-Ni correlation in low
$\afe$ field stars found by \citet{Nissen97} and in dSph galaxies
by \citet{Venn04}. Thus, to a large sense, this pair is related to
an chemical evolution history like dSph galaxies. Particularly,
this is further witnessed by the higher [Mn/Fe] than the
comparison star HD\,211998 by 0.2 dex. As noted by
\citet{Nissen00}, low $\afe$ stars in \citet{Nissen97} have higher
[Mn/Fe] than normal stars. This is consistent with their origins
from dSph galaxies because Mn is mainly produced by SN Type Ia
\citep{Samland98} which starts to contribute the chemical
evolution of dSph galaxies at lower than usual $\feh$ values in
the Galaxy.

However, the suggestion that they are accreted from dSph galaxy
may not be the whole story of this common proper-motion pair. With
careful inspection of the abundances in Table 1, some interesting
features, which are not found stars of dSph galaxies,  are shown.
The most obvious feature is that [Mg/Fe] is lower than [Ca/Fe] by
0.2 dex and than [Ti/Fe] by 0.3 dex, and [Si/Fe] is also slightly
lower than [Ca/Fe] and [Ti/Fe]. Note that the low-$\alpha$ young
globular cluster Rup 106, presumably accreted from the Sagittarius
dwarf galaxy, do not show difference between Mg-Si and Ca-Ti
abundances \citep{Brown97}. On the contrary, \citet{Venn04} found
even lower [Ca,Ti/Fe] ratios than [Mg/Fe] in metal-poor dSph
systems. Furthermore, there is a underabundant [Ba/Y] by 0.1 dex
in HD\,134439/40, which is inconsistent with most dSph stars
according to \citet{Venn04}, who found a large overabundance in
[Ba/Y] in the comparison of stars from seven dSph galaxies with
Galactic stars. It seems that additional mechanism is required to
explain the abundance pattern of this pair.

\begin{figure}
\includegraphics[width=84mm]{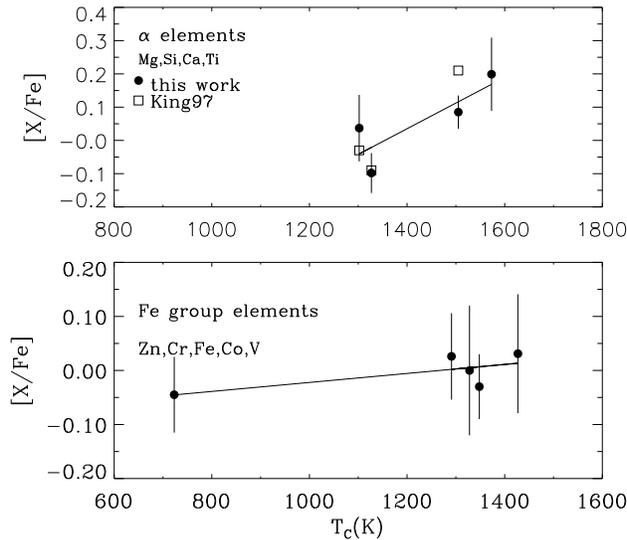}
\caption{The [X/Fe] vs. $T_{C}$ for $\alpha$ elements
and iron group elements. Elements are indicated in order of increasing $T_{C}$.}
\label{fig:AbuTcn}
\end{figure}

Since the two stars were formed in a low supernova Type II
environment as indicated by their low $\alpha$ ratios, it is
natural that dust accretion provides a possible solution to these
unexpected features. Probably, the two field low $\alpha$ stars
have formed from a dusty environment where dust from the late
stages of stellar evolution in a prior generation of stars
contaminated the pro-stellar material of this pair. In connection
with the dust observations, this suggestion is promising since the
existence of very cold dust in low metallicity galaxy has been
reported \citep{Galliano03} which indicates that the mass ratio of
dust-to-gas ratio is significantly higher than normal value in the
Galaxy \citep{Li04}. It is well known that the depletion pattern
of gas onto dust in a prior generation of stars shows a decreasing
depletion factor for elements with higher condensation
temperatures \citep{Jenkins04}. Therefore, if the proto-cloud of
this pair is polluted by material with a usually high dust-to-gas
ratio, we would expect an increasing enhancement of elements with
higher condensation temperatures. The relation of [X/Fe] vs.
$T_{C}$ for $\alpha$ elements and Fe-group elements in Fig.~5
respectively, so that small trends can be detected by comparing
elements which are thought to be formed from the same
nucleosynthesis site. In this plot, the average values of
HD\,134439 and HD\,134440 are adopted as representative abundances
of this pair. Here, Zn is included in iron group elements since it
follows Fe at $\feh \sim -1.5$ \citep{Nissen04} while Mn is
excluded due to the odd-even effect of this element found in
Galactic stars \citep{Nissen00} and Ni is also excluded due to the
Na-Ni correlation described above. As expected, a positive slope
in the [X/Fe] vs. $T_{C}$ for $\alpha$ elements is clear. This
trend is also visible based on abundances of $\alpha$ elements
from \citet{King97} as shown by open squares in Fig.~5. For Fe
group elements, the trend is generally masked within the abundance
uncertainty, but there is a hint that [Zn/Fe], being an element
least contained in dust, is lower than the average [X/Fe] of other
Fe group elements such as V and Cr. Alternatively, the hypothesis
of accreting planetesimals, which also gives rise to a positive
slope in the [X/Fe] vs. $T_{C}$, is difficult to be accepted since
there is a low probability for both stars in this common proper
motion pair have accreted the same mass of planetesimals with the
same composition. It seems that these abundances favor the dust
pollution hypothesis. Interestingly, two blue metal-poor stars
from \citet{Ivans03}, CS\,22966-043 and CS\, 22941-012, have lower
Mg and Si ratios than Ca and Ti ratios. However, it is unclear if
the dust pollution can explain \citet{Ivans03}'s result since
CS\,22966-043 is a low $\alpha$ stars while CS\, 22941-012 has
high $\afe$ (with the mean value of Mg, Si, Ca and Ti above 0.3
dex). More observational data and knowledge on dust formation and
evolution are desirable to understand these results.

\section{Conclusions}
We have derived abundances of 18 elements for the HD\,134439/40
common proper-motion pair. With new determined abundances of Mn,
Co and Zn, our result do not favor the accretion of planetesimals
suggested by \citet{Shigeyama03} and \citet{Tsujimoto06} as a
possible origin of their low $\afe$ ratios. Instead, abundance
results of low $\afe$ ratios, high [Mn/Fe] and the Na-Ni
relationship are generally consistent with the kinematical
evidence of a discrete accretion from dSph galaxy. Furthermore,
the hint of positive slopes of [X/Fe] vs. $T_{C}$ for $\alpha$
elements and probably iron group elements for this pair may
indicate that the proto-cloud of this common proper pair is
polluted by material with a usually high dust-to-gas ratio in a
low SN Type II environment such as dSph galaxy.

\section*{Acknowledgements}
Dr. Liang Yanchun is thanked
for help on data collection from UVES archive, and
Dr. Li Aigen is thanked for useful discussion on dust
observations.
This research is supported by the NSFC under grant
No. 10433010, No. 10521001 and No. 10203002, and by
the Chinese Academy of Sciences under grant No. KJCX2-SW-T06.


\begin{thebibliography}{}
\bibitem[\protect\citeauthoryear{Alonso et al.}{1996}]{Alonso96}Alonso A., Arribas S. \&  Mart\'{\i}nez-Roger C. 1996, A\&AS, 117, 227
\bibitem[\protect\citeauthoryear{Baba et al.}{2002}]{Baba02}Baba H., Yasuda N., \& Ichikama S., 2002, in Bohlender D.A., Durand D., Handley T.H., eds, ADASS XI,ASP Conference Series, Vol. 281, p.298
\bibitem[\protect\citeauthoryear{Brown et al.}{1997}]{Brown97}Brown J.A., Wallerstein G.,  Zucker D., 1997, AJ, 114, 180
\bibitem[\protect\citeauthoryear{Chen et al.}{2000}]{Chen00}Chen Y.Q., Nissen P.E., Zhao G., Benoni T., Zhang H.W., 2000, A\&AS, 141, 491
\bibitem[\protect\citeauthoryear{Chen \& Zhao}{2006}]{Chen06}Chen Y.Q., \& Zhao G., 2006, AJ, 131, 1816
\bibitem[\protect\citeauthoryear{Carney et al.}{1996}]{Carney96}Carney B.W., Laird J.B., Latham D.W., Aguilar L.A., 1996, AJ, 112, 668
\bibitem[\protect\citeauthoryear{Desidera et al.}{2004}]{Desidera04}Desidera S. , Gratton R.G., Scuderi S.  Claudi R.U., Cosentino R., Barbieri M., Bonanno G., Carretta E., Endl, M., 2004, A\&A, 420, 683
\bibitem[\protect\citeauthoryear{Fulbright}{2000}]{Fulbright00}Fulbright J.P. 2000, AJ, 120, 1841
\bibitem[\protect\citeauthoryear{Galliano et al.}{2003}]{Galliano03}Galliano F., Madden S.C., Jones A.P., Wilson C.D., Bernard J.P., Le Peintre F., 2003, A\&A, 407, 159
\bibitem[\protect\citeauthoryear{Gilmore \& Wyse}{1998}]{Gilmore98}Gilmore  G., Wyse R.F.G., 1998, AJ, 116, 748
\bibitem[\protect\citeauthoryear{Gonzalez et al.}{2001}]{Gonzalez01}Gonzalez G., Laws C., Tyagi S., Reddy, B.E., 2001, AJ, 121, 432
\bibitem[\protect\citeauthoryear{Grevesse \& Sauval}{1998}]{Grevesse98}Grevesse N., Sauval A.J., 1998, Space Science Reviews 85, 161
\bibitem[\protect\citeauthoryear{Ivans et al.}{2003}]{Ivans03}Ivans I.I, Sneden C., James C.R. , Preston G. W., Fulbright J. P., H\"oflich P. A., Carney B.W., Wheeler J. C., 2003, ApJ, 592, 906
\bibitem[\protect\citeauthoryear{Jenkins}{2004}]{Jenkins04} Jenkins E.B., 2004, in McWilliam A., Rauch  M., eds, Origin and Evolution of the Elements, Carnegie Observatories Astrophysics Series, Cambridge University Press, p.336.
\bibitem[\protect\citeauthoryear{King}{1997}]{King97}King J.R., 1997, AJ, 113, 2302
\bibitem[\protect\citeauthoryear{Kurucz}{1995}]{Kurucz95}Kurucz R.L., 1995, Atomic Line Data, Kurucz CDROM, No. 23
\bibitem[\protect\citeauthoryear{Li}{2004}]{Li04}Li A., 2004, in Block D.L., Puerari I., Freeman K., eds, Penetrating Bars Through Masks of Cosmic Dust, Springer Publisher,p.535
\bibitem[\protect\citeauthoryear{Laws \& Gonzale}{2001}]{Laws01}Laws C., Gonzalez G., 2001, ApJ, 553, 405
\bibitem[\protect\citeauthoryear{Lodders}{2003}]{Lodders03}Lodders K., 2003, ApJ, 591, 1220
\bibitem[\protect\citeauthoryear{Moore}{1966}]{Moore66}Moore C.E., Minnaert M.G.J., Houtgast J., 1966, The Solar Spectrum 2935\AA\ to 870\AA, National Bureau of Standards Monograph 61, Washington
\bibitem[\protect\citeauthoryear{Nissen et al.}{2004}]{Nissen04}Nissen P.E., Chen Y.Q., Asplund M., Pettini M., 2004, A\&A, 415, 993
\bibitem[\protect\citeauthoryear{Nissen et al.}{2000}]{Nissen00}Nissen P.E., Chen Y.Q., Schuster W.J., Zhao G., 2000, A\&A, 353, 722
\bibitem[\protect\citeauthoryear{Nissen \& Schuster}{1997}]{Nissen97}Nissen P.E. Schuster W.J., 1997, A\&A, 326, 751
\bibitem[\protect\citeauthoryear{Noguchi et al.}{2002}]{Noguchi02}Noguchi K., Aoki W., Kawanomoto S. , Ando H., Honda S., Izumiura H., Kambe E., Okita K., Sadakane K., Sato B., et al., 2002, PASJ, 54, 855
\bibitem[\protect\citeauthoryear{Sadakane et al.}{2003}]{Sadakane03}Sadakane K., Ohkubo M., Honda S., 2003, PASJ, 55, 1005
\bibitem[\protect\citeauthoryear{Samland }{1998}]{Samland98}Samland M., 1998, ApJ, 496, 155
\bibitem[\protect\citeauthoryear{Shetrone et al.}{2001}]{Shetrone01}Shetrone M.D., C\^ot\'e P., Sargent W.L., 2001, ApJ, 548, 592
\bibitem[\protect\citeauthoryear{Shetrone et al.}{2003}]{Shetrone03}Shetrone M.D., Venn K., Tolstoy E., Primas F., Hill V., Kaufer A., 2003, AJ, 125, 684.
\bibitem[\protect\citeauthoryear{Shigeyama \& Tsujimoto}{2003}]{Shigeyama03}Shigeyama T., Tsujimoto T., 2003, ApJ, 598, L47
\bibitem[\protect\citeauthoryear{Tsujimoto \& Shigeyama}{2006}]{Tsujimoto06}Tsujimoto T., Shigeyama T.,2006, ApJ, 638, L109
\bibitem[\protect\citeauthoryear{Venn et al.}{2004}]{Venn04}Venn K.A., Irwin M., Shetrone M.D., Tout C.A., Hill V., Tolstoy E., 2004, AJ, 128, 1177
\end{thebibliography}
\end{document}